\documentclass[journal,10pt]{IEEEtran}
\makeatletter
\def\endthebibliography{% 
    \def\@noitemerr{\@latex@warning{Empty `thebibliography' environment}}%
    \endlist
}
\makeatother

\usepackage{cite}

\usepackage[pdftex]{graphicx}
\usepackage[caption=false,font=footnotesize]{subfig}

\usepackage{amsmath}
\usepackage{mathtools, cuted}
\usepackage{amssymb}
\usepackage{bm}
\usepackage{mathrsfs}
\usepackage{breqn}

\usepackage{pifont}
\usepackage{xcolor}
\usepackage{url}
\usepackage{lettrine}
\usepackage{lipsum}
\usepackage{siunitx}
\usepackage{soul}
\usepackage{array}
\usepackage[inline]{enumitem}
\usepackage{epsfig}
\usepackage{filecontents}

\usepackage{algpseudocode}
\usepackage{algorithm, tabularx}
\usepackage{multirow}
\newcolumntype{L}[1]{>{\raggedright\let\newline\\\arraybackslash\hspace{0pt}}m{#1}}
\newcolumntype{C}[1]{>{\centering\let\newline\\\arraybackslash\hspace{0pt}}m{#1}}
\newcolumntype{R}[1]{>{\raggedleft\let\newline\\\arraybackslash\hspace{0pt}}m{#1}}
\newlength{\maxwidth}

\makeatletter
\newcommand{\multiline}[1]{%
	\begin{tabularx}{\dimexpr\linewidth-\ALG@thistlm}[t]{@{}X@{}}
		#1
	\end{tabularx}
}
\makeatother
\algdef{SE}[SUBALG]{Indent}{EndIndent}{}{{\algorithmicend\ }}
\algtext*{Indent}
\algtext*{EndIndent}

\usepackage{amsthm}

\theoremstyle{remark}

\begin{document}

\title{\LARGE Enabling Secure Wireless Communications for FARIS-Aided Systems}

\author{Hong-Bae Jeon,~\IEEEmembership{Member,~IEEE}, Yonghwi Kim,~\IEEEmembership{Member,~IEEE,} Hyung-Joo Moon,~\IEEEmembership{Member,~IEEE,} and\\Kai-Kit Wong,~\IEEEmembership{Fellow,~IEEE} %and Chan-Byoung Chae,~\IEEEmembership{Fellow,~IEEE}%
\thanks{\textit{(Corresponding Author: Hyung-Joo Moon)}}%
\thanks{H.-B. Jeon is with the School of Electronic Engineering, Soongsil University, Seoul, Korea (e-mail: hongbae08@ssu.ac.kr).}%
\thanks{Y. Kim is with the Department of Electronics and Electrical Engineering, Dankook University, Yongin, Korea (e-mail: eric\_kim@dankook.ac.kr).}
\thanks{H.-J. Moon is with the School of Integrated Technology, Yonsei University, Seoul, Korea (e-mail: moonhj@yonsei.ac.kr).}%
	\thanks{K.-K. Wong is with the Department of Electronic and Electrical Engineering, University College London, WC1E 6BT London, U.K., and also with the Yonsei Frontier Laboratory, Yonsei University, Seoul, Korea (e-mail: kai-kit.wong@ucl.ac.uk).}
%\thanks{C.-B. Chae is with the School of Integrated Technology, Yonsei University, Seoul 03722, Korea (e-mail: cbchae@yonsei.ac.kr).}%
}

\maketitle

\begin{abstract}
This paper investigates secure downlink transmission assisted by a fluid active reconfigurable intelligent surface (FARIS), which enables both active reflection and dynamic port selection, offering enhanced flexibility for physical-layer security. We formulate a secrecy rate maximization problem that jointly optimizes the transmit beamformer, active reflection coefficients, and fluid port configuration under practical power constraints. To efficiently handle the resulting highly nonconvex problem, we develop a tailored alternating optimization (AO) framework that decomposes the original joint design into tractable subproblems, where each admits an efficient solution while preserving the system constraints, enabling an effective joint optimization of beamforming and FARIS reconfiguration. Numerical results demonstrate that the proposed FARIS-assisted design consistently outperforms the benchmarks. The results further highlight the robustness of FARIS against unfavorable eavesdropping geometries, confirming its potential as a powerful enabler for secure communications in challenging environments.
\end{abstract}


\IEEEpeerreviewmaketitle

\section{Introduction}
\label{sec:intro}
The broadcast nature of wireless communications inherently exposes transmitted signals to eavesdropping, making security a fundamental challenge. Physical-layer security (PLS) has emerged as a promising approach to safeguard confidential information by exploiting the randomness and spatial characteristics of wireless channels, without relying on higher-layer cryptographic mechanisms~\cite{pls}. By judiciously designing transmission strategies to enhance the legitimate channel while suppressing information leakage, PLS offers an information-theoretic complement to conventional security techniques~\cite{rispls}.

Recently, reconfigurable intelligent surface (RIS) have attracted considerable attention as a disruptive technology for shaping the wireless propagation environment~\cite{risspm}. By intelligently adjusting the phase responses of a large number of low-cost reflecting elements, RIS enables favorable signal manipulation that can enhance coverage, spectral efficiency, and secrecy performance~\cite{HBRIS, nfris}. In the context of PLS, RIS can reinforce the legitimate signal and degrade the eavesdropper’s reception by exploiting passive beamforming~\cite{rispls}. However, conventional RIS relies on passive elements with fixed locations, which fundamentally limits their capability to fully adapt to dynamic and adversarial wireless environments.

To overcome the rigidity of fixed-position RIS, the concept of fluid-RIS (FRIS) has been introduced~\cite{FRISmag}, motivated by fluid antenna~\cite{fas} where reflecting elements can be dynamically selected from candidate locations. By enabling position reconfigurability, FRIS unlocks an additional spatial degree-of-freedom (DoF) beyond phase control, allowing the system to escape unfavorable channel realizations and achieve enhanced secrecy performance~\cite{FRISsec, FRISsec22}. Despite these advantages, existing FRIS designs remain constrained by their passive nature, which restricts the achievable signal enhancement and limits robustness in scenarios with severe path loss or blockage.

Motivated by these, this paper considers a more general and powerful paradigm; the fluid-active-RIS (FARIS), first introduced by \textit{\textbf{Jeon}}~\cite{FARIS}. Unlike conventional architectures, FARIS integrates active reflection with dynamic port selection, enabling both controllable amplification and spatial reconfiguration. This hybrid capability allows FARIS to not only reshape the wireless environment but also compensate for path loss and improve resilience against adverse channel conditions.

From a PLS perspective, the signal amplification and additional DoF enabled by FARIS open up new opportunities for secrecy enhancement by simultaneously strengthening the legitimate channel while suppressing information leakage to eavesdroppers. Motivated by this, this paper investigates a FARIS-assisted system in the presence of an eavesdropper, where the secrecy rate is maximized through the joint optimization of transmit beamforming and FARIS configurations. The main contributions are summarized as follows:
\begin{itemize}
\item We formulate a secrecy rate maximization problem for the FARIS-assisted system, jointly optimizing the transmit beamformer, the selection of FARIS ports, and their reflection and amplification coefficients, while accounting for transmit and radiated power limits. 
\item Due to its coupled continuous and combinatorial nature, we develop an alternating optimization (AO) framework that decomposes it into tractable subproblems. By leveraging closed-form beamforming updates and selection strategies, the proposed design effectively exploits the joint benefits of active reflection and fluid port selection.
\item Numerical results demonstrate that the proposed FARIS-based scheme consistently outperforms the benchmarks, highlighting the substantial secrecy gains by FARIS.
\end{itemize}
\begin{figure}[t]
  \begin{center}
    \includegraphics[width=0.95\columnwidth,keepaspectratio]{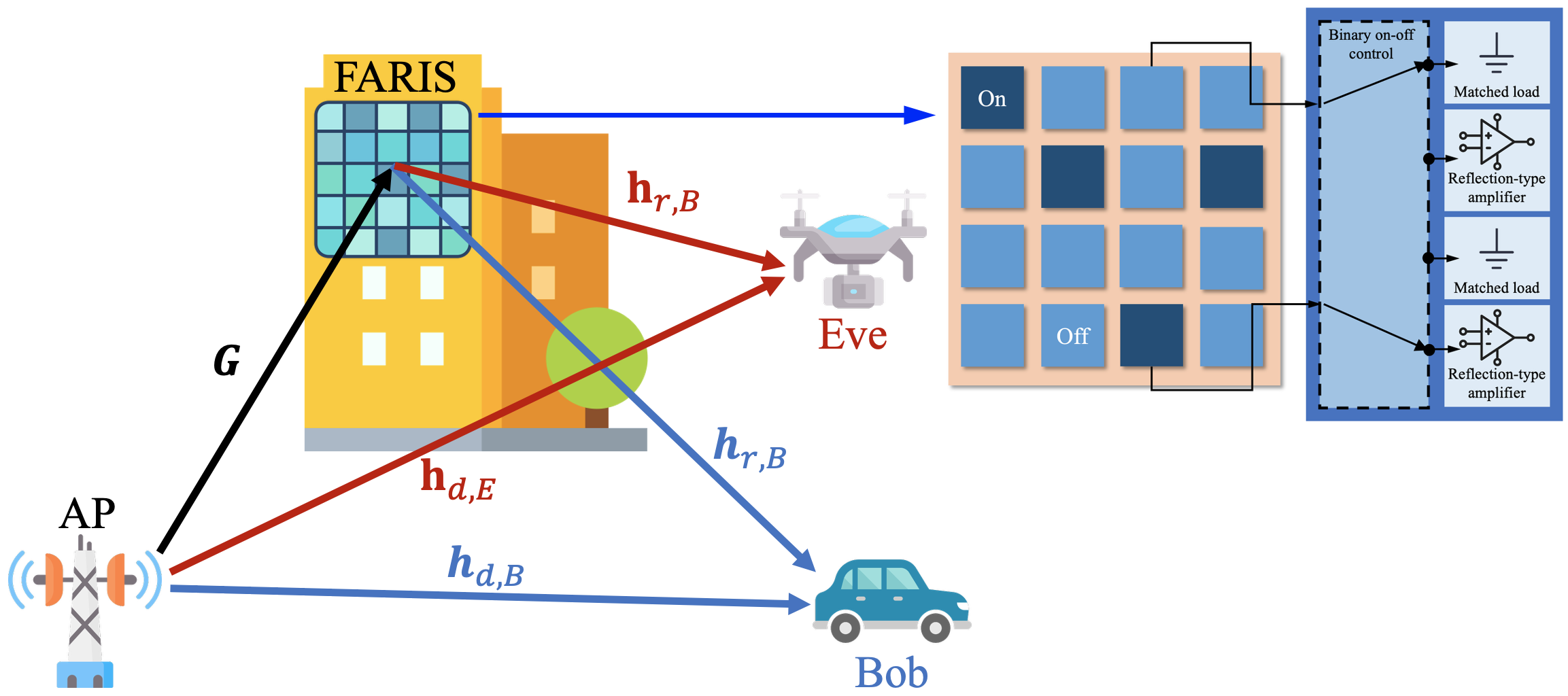}
    \caption{FARIS-assisted secure communication system with the illustration of FARIS schematic.}
    \label{fig_sys}
  \end{center}
\end{figure}
\section{System Model}
\label{sec:sys}
As depicted in Fig.~\ref{fig_sys}, we consider a secure downlink system, where an $N$-antenna access point (AP) communicates with a single-antenna legitimate user (Bob) in the presence of a single-antenna eavesdropper (Eve). To enhance PLS, an $M=M_x \times M_x$-element FARIS is deployed, whose reflective elements uniformly distributed over $W_x\lambda \times W_x\lambda$-size square surface, where $\lambda$ denotes the carrier wavelength and $W_x$ represents the $\lambda$-normalized aperture size. The resulting inter-element spacing is given by $d=\frac{W_x\lambda}{M_x}$. Due to the finite spacing between elements, spatial correlation naturally arises across the FARIS aperture. This effect is captured by a correlation matrix $\mathbf J\in\mathbb R^{M\times M}$, whose $ij$ entry is modeled using Jakes' correlation model as $J_{ij}=j_0\left(\frac{2\pi d_{ij}}{\lambda}\right)$~\cite{FRISonoff}, where $j_0(\cdot)$ denotes the zero-order spherical Bessel function of the first kind, and $d_{ij}$ represents the distance between the $i$ and $j$th elements. Following~\cite{FRISonoff, FRISmag}\footnote{Although the referenced model is developed for FRIS, it is adopted to FARIS~\cite{FARIS} since it shares the same underlying hardware architecture, with the only difference being the additional amplification module.} and demonstrated in Fig.~\ref{fig_sys}, each FARIS element is interpreted as a port of a fluid antenna structure~\cite{FARIS}. Each port operates in one of two modes; In the \textbf{on} state, the port actively interacts with the incident electromagnetic wave and applies controllable amplitude and phase modifications. In the \textbf{off} state, the port is terminated with a matched load, effectively isolating it from the impinging signal and preventing any reflection. 

At each transmission block, only $M_o< M$ ports are activated. Let $\mathbf S_{M_o}^{\mathrm{T}}\in\{0,1\}^{M\times M_o}$ denote the port-selection matrix whose columns are mutually distinct canonical basis:
\begin{equation}
\label{smo}
\mathbf S_{M_o}^{\mathrm T} = [\mathbf e_{i_1} \cdots \mathbf e_{i_{M_o}}]~(\forall i_m\in\{1,\cdots,M\}),
\end{equation}
where $\mathbf e_i\in\mathbb R^M$ denotes the canonical basis vector; the $i$th component is 1 and others are zero. 
Let $\mathbf v=[v_1 \cdots v_{M_o}]^{\mathrm T}\in\mathbb C^{M_o}$ denote the active reflection coefficients with $v_i = g_i e^{j\phi_i},~ g_i\in[0,g_{\max}], \phi_i\in[0,2\pi)$. The FARIS effective operator is defined as $\mathbf A_{\rm act}
\triangleq
\mathbf J^{1/2}\mathbf S_{M_o}^{\mathrm T}
\mathrm{diag}(\mathbf v)
\mathbf S_{M_o}\mathbf J^{1/2}$. 

Let $\mathbf h_{d,B},\mathbf h_{d,E}\in\mathbb C^{N\times 1}$ denote the direct AP-Bob and AP-Eve channels, respectively. Let $\mathbf G\in\mathbb C^{M\times N}$ denote the AP-FARIS channel, and $\mathbf h_{r,B},\mathbf h_{r,E}\in\mathbb C^{M\times 1}$ denote the FARIS-Bob and FARIS-Eve channels, respectively. The AP transmits $x$ with $\mathbb E[|x|^2]=1$ using a beamformer $\mathbf w\in\mathbb C^{N\times 1}$ subject to $\|\mathbf w\|_2^2\le P_{\rm AP}$.
Since FARIS is active, its internal thermal noise is denoted by $\mathbf n_r\sim\mathcal{CN}(\mathbf 0,\sigma_r^2\mathbf I_M)$.
The signal radiated by FARIS is therefore given by
\begin{equation}
\label{sris}
\begin{aligned}
&\mathbf s_{\rm RIS}
=
\mathbf A_{\rm act}\big(\mathbf G\mathbf w x+\mathbf n_r\big),\\
&\rightarrow\mathbb E[\|\mathbf s_{\rm RIS}\|_2^2]
=
\|\mathbf A_{\rm act}\mathbf G\mathbf w\|_2^2
+
\sigma_r^2 \mathrm{tr}\left(\mathbf A_{\rm act}\mathbf A_{\rm act}^{*}\right).
\end{aligned}
\end{equation}
In addition to $\mathbb E[\|\mathbf s_{\rm RIS}\|_2^2]$, the operation of FARIS also incurs hardware power consumption associated with the selected fluid-active ports. Specifically, based on the hardware architecture and circuit model in Fig.~\ref{fig_sys}, the circuit power consumption of FARIS consists of two components: i) $P_c$, representing the logical control and switching power consumed by each candidate element, and ii) $P{\mathrm{DC}}$, denoting the direct current (DC) bias power required for active reflection~\cite{aris5}. Since the logical control circuitry is required for all $M$ candidate elements, while the active reflection branch is activated only for the selected $M_o$ FARIS ports, the total FARIS power consumption is modeled as
\begin{equation}
\label{eq:Pfaris_total}
P_{\mathrm{FARIS}}^{\mathrm{tot}}
\triangleq
MP_c + M_oP_{\mathrm{DC}} + \xi\mathbb E[\|\mathbf s_{\rm RIS}\|_2^2],\footnote{Note that when $M_o=M$, the above expression reduces to the conventional ARIS power consumption model in~\cite{aris5}.}
\end{equation}
where $\xi \triangleq \frac{1}{\upsilon}$ and $\upsilon\in(0,1]$ denotes the amplifier efficiency. Accordingly, under the total FARIS power budget $P_{\max,t}$, the feasible reflection design must satisfy $P_{\mathrm{FARIS}}^{\mathrm{tot}} \le P_{\max,t}$:%, which is equivalent to
\begin{equation}
\label{eq:Pris_constraint}
\mathbb E[\|\mathbf s_{\rm RIS}\|_2^2]
\le
\upsilon\big(P_{\max,t}-MP_c-M_oP_{\mathrm{DC}}\big)
\triangleq
P_{\max}.
\end{equation}
The received signals at Bob and Eve are hence given by $
y_k=(\mathbf h_{d,k}^{*}+\mathbf h_{r,k}^{*}\mathbf A_{\rm act}\mathbf G)\mathbf w x
+
\mathbf h_{r,k}^{*}\mathbf A_{\rm act}\mathbf n_r
+
z_k$, where $k\in\{B,E\}$ and $z_k\sim\mathcal{CN}(0,\sigma_0^2)$ are noise at receiver.

Define the effective channels of Bob and Eve, respectively: $
%h_B \triangleq \mathbf h_{d,B}+\mathbf G^{*}\mathbf A_{\rm act}^{*}\mathbf h_{r,B},~
h_k \triangleq \mathbf h_{d,k}+\mathbf G^{*}\mathbf A_{\rm act}^{*}\mathbf h_{r,k}$. Thereby, the signal-to-interference-noise-ratio (SINR) at Bob and Eve are hence:
\begin{equation}
\label{sinrbe}
\gamma_k
=
\frac{|h_k^{*}\mathbf w|^2}
{\sigma_0^2+\sigma_r^2 \mathbf h_{r,k}^{*}\mathbf A_{\rm act}\mathbf A_{\rm act}^{*}\mathbf h_{r,k}}.
\end{equation}
The achievable secrecy rate is defined as~\cite{FRISsec}
\begin{equation}
R_s
=
\big[\log_2(1+\gamma_B)-\log_2(1+\gamma_E)\big]^+~([x]^+\triangleq \max(x, 0)).
\end{equation}
Thus, on the positive-secrecy region $\{\gamma_B\ge\gamma_E\}$, maximizing $R_s$ is equivalent to maximizing $\frac{1+\gamma_B}{1+\gamma_E}$.

Our objective is to maximize $R_s$, which is achieved by jointly optimizing $\mathbf w,~\mathbf S_{M_o}$, and $\mathbf v$. Motivated by this, we consider the secrecy-ratio maximization problem
\begin{equation}\label{prob:P1}
\begin{aligned}
&\max_{\mathbf w,\mathbf v,\mathbf S_{M_o}}~
\frac{1+\gamma_B(\mathbf w,\mathbf v,\mathbf S_{M_o})}
{1+\gamma_E(\mathbf w,\mathbf v,\mathbf S_{M_o})}\\
\text{s.t.}~
&\|\mathbf w\|_2^2\le P_{\rm AP}, \forall|v_i|\le g_{\max},~\eqref{smo},~\eqref{eq:Pris_constraint}.
\end{aligned}
\end{equation}
Problem~\eqref{prob:P1} is a mixed-integer nonlinear program (MINLP) due to the fractional secrecy objective, the coupling among $(\mathbf w,\mathbf v,\mathbf S_{M_o})$, and the combinatorial port selection. We develop an efficient algorithm based on an AO framework. In each AO iteration, we alternate between optimizing $\mathbf w$, $\mathbf v$, and $\mathbf S_{M_o}$.

\section{Proposed AO Framework}
\label{sec:ao}
\subsection{Optimization of $\mathbf w$: Transmit Beamforming}
\label{subsec:w_opt}

For fixed $(\mathbf v,\mathbf S_{M_o})$, $\{h_k\}$ and the effective noise powers $\eta_k=\sigma_0^2+\sigma_r^2\mathbf h_{r,k}^{*}\mathbf A_{\rm act}\mathbf A_{\rm act}^{*}\mathbf h_{r,k}$ are fixed as well. Define $\mathbf H_k \triangleq h_k h_k^{*}\succeq \mathbf 0,~\mathbf B \triangleq \mathbf G^{*}\mathbf A_{\rm act}^{*}\mathbf A_{\rm act}\mathbf G \succeq \mathbf 0$, and $\bar P_{\rm RIS}\triangleq P_{\max}-\sigma_r^2\operatorname{tr}\left(\mathbf A_{\rm act}\mathbf A_{\rm act}^{*}\right)$. Then,~\eqref{prob:P1} becomes
\begin{equation}\label{prob:P2_exact}
\begin{aligned}
\max_{\mathbf w}~ 
\frac{\eta_B+\mathbf w^{*}\mathbf H_B\mathbf w}
{\eta_E+\mathbf w^{*}\mathbf H_E\mathbf w}~\text{s.t.}~
\mathbf w^{*}\mathbf w\le P_{\rm AP}, \mathbf w^{*}\mathbf B\mathbf w\le \bar P_{\rm RIS}.
\end{aligned}
\end{equation}
To solve~\eqref{prob:P2_exact}, we introduce the lifted variable $\mathbf W \triangleq \mathbf w\mathbf w^{*}\succeq \mathbf 0~(\operatorname{rank}(\mathbf W)=1)$. Then,~\eqref{prob:P2_exact} can be rewritten as
\begin{equation}\label{prob:P2_W}
\begin{aligned}
&\max_{\mathbf W\succeq \mathbf 0}~\frac{\eta_B+\operatorname{tr}(\mathbf H_B\mathbf W)}
{\eta_E+\operatorname{tr}(\mathbf H_E\mathbf W)}\\
&\text{s.t.}~\operatorname{tr}(\mathbf W)\le P_{\rm AP}, \operatorname{tr}(\mathbf B\mathbf W)\le \bar P_{\rm RIS}, \operatorname{rank}(\mathbf W)=1.
\end{aligned}
\end{equation}
Next, applying the Charnes-Cooper transformation~\cite{charnes} with $\tau \triangleq \frac{1}{\eta_E+\operatorname{tr}(\mathbf H_E\mathbf W)}$ and $\mathbf X \triangleq \tau\mathbf W$, we equivalently obtain:
\begin{equation}\label{prob:P2_CC_rank}
\begin{aligned}
&\max_{\mathbf X\succeq\mathbf 0, \tau\ge 0}~
\tau\eta_B+\operatorname{tr}(\mathbf H_B\mathbf X)\\
\text{s.t.}~
&\tau\eta_E+\operatorname{tr}(\mathbf H_E\mathbf X)=1, \operatorname{tr}(\mathbf X)\le \tau P_{\rm AP},\\
&\operatorname{tr}(\mathbf B\mathbf X)\le \tau \bar P_{\rm RIS}, \operatorname{rank}(\mathbf X)=1.
\end{aligned}
\end{equation}
By dropping the rank-one constraint from~\eqref{prob:P2_CC_rank}, we arrive at the semidefinite relaxation, which is a convex semidefinite programming (SDP) and can be efficiently solved by CVX~\cite{boyd}. After obtaining the optimal $(\mathbf X^\star,\tau^\star)$ with $\mathbf W^\star=\frac{\mathbf X^\star}{\tau^\star}$, Gaussian randomization is applied to recover a feasible $\mathbf w^\star$~\cite{sdr}.

\subsection{Update of $\mathbf v$: Reflection and Amplification Coefficients}
\label{subsubsec:v_update_full_direct}
For fixed $(\mathbf w, \mathbf S_{M_o})$, define $\mathbf T \triangleq \mathbf S_{M_o}\mathbf J^{1/2}\in\mathbb C^{M_o\times M},~\mathbf K \triangleq \mathbf T\mathbf T^{*}=\mathbf S_{M_o}\mathbf J\mathbf S_{M_o}^{\mathrm T}\succeq\mathbf 0$. Then, $\mathbf A_{\rm act}$ can be written as $\mathbf A_{\rm act}=\mathbf T^{*}\mathrm{diag}(\mathbf v)\mathbf T$. Now define $c_k \triangleq \mathbf h_{d,k}^{*}\mathbf w,~\mathbf b \triangleq \mathbf T\mathbf G\mathbf w\in\mathbb C^{M_o}$, and $\mathbf u_k \triangleq \mathbf T\mathbf h_{r,k}\in\mathbb C^{M_o}$. Then the FARIS-assisted terms satisfy $\mathbf h_{r,k}^{*}\mathbf A_{\rm act}\mathbf G\mathbf w=\mathbf a_k^{\mathrm{T}}\mathbf v$, where $\mathbf a_k \triangleq \mathbf u_k^\ast\odot \mathbf b$. Then $h_k^{*}\mathbf w=c_k+\mathbf a_k^{\mathrm T}\mathbf v$ holds. Thus, we have the numerator in~\eqref{sinrbe}
\begin{equation}
\label{num5}
\big|c_k+\mathbf a_k^{\mathrm T}\mathbf v\big|^2=|c_k|^2+2\Re\{\mathbf d_k^{*}\mathbf v\}+\mathrm{tr}(\mathbf A_k\mathbf V),
\end{equation}
where $\mathbf d_k \triangleq c_k\mathbf a_k^\ast\in\mathbb C^{M_o}, \mathbf A_k \triangleq \mathbf a_k^\ast\mathbf a_k^{\mathrm T}\succeq\mathbf 0$, and $\mathbf V\triangleq \mathbf v\mathbf v^{*}\succeq\mathbf 0~ (\mathrm{rank}(\mathbf V)=1)$. For the denominator in~\eqref{sinrbe}, using $\mathrm{diag}(\mathbf v)\mathbf K\mathrm{diag}(\mathbf v^\ast)=\mathbf K\odot(\mathbf v\mathbf v^{*})$:
\begin{equation}
\sigma_r^2\mathbf h_{r,k}^{*}\mathbf A_{\rm act}\mathbf A_{\rm act}^{*}\mathbf h_{r,k}=\sigma_r^2\mathbf u_k^{*}\mathrm{diag}(\mathbf v)\mathbf K\mathrm{diag}(\mathbf v^\ast)\mathbf u_k
%&=\sigma_r^2 \mathrm{tr}\Big(\big(\mathbf K\odot(\mathbf u_k\mathbf u_k^{*})\big)(\mathbf v\mathbf v^{*})\Big)\\
=\mathrm{tr}(\mathbf C_k \mathbf V),
\label{eq:ampnoise_full}
\end{equation}
where $\mathbf C_k \triangleq \sigma_r^2\big(\mathbf K\odot(\mathbf u_k\mathbf u_k^{*})\big)\succeq\mathbf 0$. Hence,~\eqref{sinrbe} becomes
\begin{equation}
\gamma_k(\mathbf v,\mathbf V)
=
\frac{|c_k|^2+2\Re\{\mathbf d_k^{*}\mathbf v\}+\mathrm{tr}(\mathbf A_k\mathbf V)}
{\sigma_0^2+\mathrm{tr}(\mathbf C_k\mathbf V)}.
\label{eq:sinr_full_inclusion}
\end{equation}
For~\eqref{sris}, the radiated signal power can be expressed in terms of $\mathbf v$. Using $\mathbf A_{\rm act}=\mathbf T^{*}\mathrm{diag}(\mathbf v)\mathbf T$ and $\mathbf b \triangleq \mathbf T\mathbf G\mathbf w$, we obtain
%HJM 아래 equation에서 diag(b)*Kdiag(b)V가 맞는것 같아 확인 부탁드립니다. 그 아래 F(w)에 대한 정의와 
\begin{equation}
\label{nof}
%\begin{aligned}
\|\mathbf A_{\rm act}\mathbf G\mathbf w\|_2^2
%&=
%\mathbf b^{*}\mathrm{diag}(\mathbf v)\mathbf K
%\mathrm{diag}(\mathbf v)^{*}\mathbf b\\
=
\mathrm{tr}\left(
\mathrm{diag}(\mathbf b)^{*}\mathbf K
\mathrm{diag}(\mathbf b)\mathbf V
\right),
%\end{aligned}
\end{equation}
where $\mathbf V=\mathbf v\mathbf v^{*}\succeq\mathbf 0$. Next, $\mathbf A_{\rm act}\mathbf A_{\rm act}^{*}
=
\mathbf T^{*}\mathrm{diag}(\mathbf v)
\mathbf K
\mathrm{diag}(\mathbf v)^{*}\mathbf T$ holds, which yields $\mathrm{tr}(\mathbf A_{\rm act}\mathbf A_{\rm act}^{*})
=
\mathrm{tr}\left(
\mathrm{diag}(\mathbf v)\mathbf K
\mathrm{diag}(\mathbf v)^{*}\mathbf K
\right)$. Using the identity
$
\mathrm{diag}(\mathbf v)\mathbf K\mathrm{diag}(\mathbf v)^{*}
=
\mathbf K\odot\mathbf V,
$
we further obtain
\begin{equation}
\label{tra2}
\mathrm{tr}(\mathbf A_{\rm act}\mathbf A_{\rm act}^{*})
=
\mathrm{tr}\left(
(\mathbf K\odot\mathbf K^{\mathrm T})\mathbf V
\right).
\end{equation}
Therefore,~\eqref{sris} can be written as $\mathbb E[\|\mathbf s_{\rm RIS}\|_2^2]=
\mathrm{tr}\left(
\mathbf F(\mathbf w)\mathbf V
\right)$, where $\mathbf F(\mathbf w)
\triangleq
\mathrm{diag}(\mathbf b)^{*}\mathbf K
\mathrm{diag}(\mathbf b)
+
\sigma_r^2(\mathbf K\odot\mathbf K^{\mathrm T})$. 
%For~\eqref{sris}, we obtain $\|\mathbf A_{\rm act}\mathbf G\mathbf w\|_2^2=\mathrm{tr}\Big(\mathrm{diag}(\mathbf b)\mathbf K\mathrm{diag}(\mathbf b)^{*}\mathbf V\Big)$, and $\mathrm{tr}(\mathbf A_{\rm act}\mathbf A_{\rm act}^{*})
%=
%\mathrm{tr}\Big(\mathbf K\mathrm{diag}(|\mathbf v|^2)\Big)
%=
%\mathrm{tr}\Big(\mathrm{diag}(\mathrm{diag}(\mathbf K))\mathbf V\Big)$. Therefore,~\eqref{sris} becomes
%\begin{equation}
%\begin{aligned}
%&\mathbb E[\|\mathbf s_{\rm RIS}\|_2^2]=\mathrm{tr}\Big(\mathbf F(\mathbf w)\mathbf V\Big)\le P_{\max},\\
%&\mathbf F(\mathbf w)\triangleq
%\mathrm{diag}(\mathbf b)\mathbf K\mathrm{diag}(\mathbf b)^{*}
%+
%\sigma_r^2\mathrm{diag}(\mathrm{diag}(\mathbf K)),
%\end{aligned}
%\label{eq:F_full}
%\end{equation}
Hence,~\eqref{prob:P1} with respect to $\mathbf v$ and $\mathbf V$ becomes:
\begin{equation}\label{prob:P3_full}
\begin{aligned}
&\max_{\mathbf v,\mathbf V\succeq\mathbf 0}~ \frac{1+\gamma_B(\mathbf v,\mathbf V)}{1+\gamma_E(\mathbf v,\mathbf V)}\\
\text{s.t.}~
&\mathrm{tr}(\mathbf F(\mathbf w)\mathbf V)\le P_{\max}, \forall [\mathbf V]_{ii}\le g_{\max}^2, \mathbf V=\mathbf v\mathbf v^{*}.
\end{aligned}
\end{equation}
Problem~\eqref{prob:P3_full} is still nonconvex due to the fractional objective structure and the rank-one coupling constraint $\mathbf V=\mathbf v\mathbf v^{*}$. To convexify~\eqref{prob:P3_full}, we start from~\eqref{eq:sinr_full_inclusion}. To obtain a closed-form lifted representation that absorbs the linear cross-term, we introduce the augmented variable $\mathbf X \triangleq
\begin{bmatrix}
\mathbf V & \mathbf v\\
\mathbf v^{*} & 1
\end{bmatrix}\succeq \mathbf 0$. Thereafter, define the augmented matrix $\mathbf Q_k \triangleq
\begin{bmatrix}
\mathbf A_k & \mathbf d_k\\
\mathbf d_k^{*} & |c_k|^2
\end{bmatrix}
=
\begin{bmatrix}\mathbf a_k^\ast\\ c_k^\ast\end{bmatrix}
\begin{bmatrix}\mathbf a_k^{\mathrm T} & c_k\end{bmatrix}
\succeq\mathbf 0$, which makes $|c_k+\mathbf a_k^{\mathrm T}\mathbf v|^2=\mathrm{tr}(\mathbf Q_k\mathbf X)$. Now define $\tilde{\mathbf C}_k \triangleq
\begin{bmatrix}
\mathbf C_k & \mathbf 0\\
\mathbf 0^{*} & 0
\end{bmatrix}\succeq\mathbf 0$, so that $\mathrm{tr}(\mathbf C_k\mathbf V)=\mathrm{tr}(\tilde{\mathbf C}_k\mathbf X)$. Thereby,~\eqref{eq:sinr_full_inclusion} becomes $\gamma_k(\mathbf X)
=
\frac{\mathrm{tr}(\mathbf Q_k\mathbf X)}
{\sigma_0^2+\mathrm{tr}(\tilde{\mathbf C}_k\mathbf X)}$. Thereafter, $\forall|v_i|\le g_{\max}$ is enforced by $\forall[\mathbf X]_{ii}=[\mathbf V]_{ii}\le g_{\max}^2$, and $\mathrm{tr}(\mathbf F(\mathbf w)\mathbf V)\le P_{\max}$ can be written as $
\mathrm{tr}(\tilde{\mathbf F}(\mathbf w)\mathbf X)\le P_{\max}$, where $\tilde{\mathbf F}(\mathbf w)\triangleq
\begin{bmatrix}
\mathbf F(\mathbf w) & \mathbf 0\\
\mathbf 0^{*} & 0
\end{bmatrix}\succeq\mathbf 0$. Hence,~\eqref{prob:P3_full} with dropping $\mathbf V=\mathbf v\mathbf v^{*}$ becomes
\begin{equation}\label{prob:P3X}
\begin{aligned}
&\max_{\mathbf X\succeq\mathbf 0}~
 \frac{1+\gamma_B(\mathbf X)}{1+\gamma_E(\mathbf X)}
\\
&\text{s.t.}~
[\mathbf X]_{M_o+1,M_o+1}=1, \mathrm{tr}(\tilde{\mathbf F}(\mathbf w)\mathbf X)\le P_{\max}, \forall[\mathbf X]_{ii}\le g_{\max}^2.
\end{aligned}
\end{equation}
Problem~\eqref{prob:P3X} is still nonconvex due to the fractional objective. To solve this, define the following affine functions of $\mathbf X$: $u_k(\mathbf X)\triangleq \mathrm{tr}(\mathbf Q_k\mathbf X)\ge 0,~v_k(\mathbf X)\triangleq \sigma_0^2+\mathrm{tr}(\tilde{\mathbf C}_k\mathbf X)>0$. Then $\gamma_k(\mathbf X)=\frac{u_k(\mathbf X)}{v_k(\mathbf X)}$, and by the quadratic transform~\cite{fracp}:
\begin{equation}
\frac{u_k(\mathbf X)}{v_k(\mathbf X)}
=
\max_{y_k\ge0} \Big(2y_k\sqrt{u_k(\mathbf X)}-y_k^2 v_k(\mathbf X)\Big),
\label{eq:quad_transform2}
\end{equation}
and the maximizer is $y_k^\star=\frac{\sqrt{u_k(\mathbf X)}}{v_k(\mathbf X)}$. %Equation~\eqref{eq:quad_transform2} converts a ratio into a concave quadratic form with respect to $y_k\ge 0$ and yields a closed-form update for $y_k$ given the other variables. 
Thereafter, we equivalently maximize $\log_2\big(1+\gamma_B(\mathbf X)\big)-\log_2\big(1+\gamma_E(\mathbf X)\big)$, which is monotone with respect to the original ratio objective. For fixed $\{y_k\}$, define the surrogate SINR function $\hat\gamma_k(\mathbf X)\triangleq
2y_k\sqrt{u_k(\mathbf X)}-y_k^2v_k(\mathbf X)$. Substituting this expression yields:%into the secrecy objective yields
\begin{equation}
\label{surop}
\max_{\mathbf X\succeq0}~
\log_2(1+\hat\gamma_B(\mathbf X))
-
\log_2(1+\hat\gamma_E(\mathbf X))~\mathrm{s.t.}~\mathrm{constraints~in~\eqref{prob:P3X}}.
\end{equation}
The objective of~\eqref{surop} is still nonconcave due to its difference-of-concave structure. To obtain a tractable surrogate, %we construct a concave minorizer of the secrecy objective. 
 for fixed $y_E$, define $g_E(\mathbf X)\triangleq \log_2\bigl(1+\hat\gamma_E(\mathbf X)\bigr)$. %Since \(\hat\gamma_E(\mathbf X)=2y_E\sqrt{u_E(\mathbf X)}-y_E^2v_E(\mathbf X)\) is concave in \(\mathbf X\), and \(\log_2(1+x)\) is concave and nondecreasing over its domain, \(g_E(\mathbf X)\) is also concave. Therefore, 
Its first-order expansion at current \(\mathbf X^{(i)}\) provides the following affine upper-bound: %For \(u_E^{(i)}>0\), % and \(1+\hat\gamma_E^{(i)}>0\),
\begin{equation}
\begin{aligned}
%&g_E(\mathbf X)\\
%&\le\bar g_E^{(i)}(\mathbf X)
\bar g_E^{(i)}(\mathbf X)\triangleq&
\log_2\bigl(1+\hat\gamma_E^{(i)}\bigr)+
\frac{1}
{\bigl(1+\hat\gamma_E^{(i)}\bigr)\ln 2}\\
&\Bigg[
\frac{y_E}{\sqrt{u_E^{(i)}}}
\bigl(u_E(\mathbf X)-u_E^{(i)}\bigr)-y_E^2
\bigl(v_E(\mathbf X)-v_E^{(i)}\bigr)
\Bigg],
\end{aligned}
\label{eq:mm_upper}
\end{equation}
where $u_E^{(i)}
\triangleq
u_E(\mathbf X^{(i)})>0$, $v_E^{(i)}
\triangleq
v_E(\mathbf X^{(i)})$, and corresponding $\hat\gamma_E^{(i)}
\triangleq
2y_E\sqrt{u_E^{(i)}}-y_E^2v_E^{(i)}
=
\hat\gamma_E(\mathbf X^{(i)})$.
%Since \(u_E(\mathbf X)\) and \(v_E(\mathbf X)\) are affine in \(\mathbf X\), so is \(\bar g_E^{(i)}(\mathbf X)\).
%Equivalently, the non-constant part of $\bar g_E^{(i)}(\mathbf X)$ can be expressed:
%\begin{equation}
%\label{noncdgg}
%\frac{
%\frac{y_E}{\sqrt{u_E^{(i)}}}
%\operatorname{tr}\left(
%\mathbf Q_E(\mathbf X-\mathbf X^{(i)})
%\right)
%-y_E^2
%\operatorname{tr}\left(
%\tilde{\mathbf C}_E(\mathbf X-\mathbf X^{(i)})
%\right)}
%{\bigl(1+\hat\gamma_E^{(i)}\bigr)\ln 2}.
%\end{equation}
Using~\eqref{eq:mm_upper} %as an upper-bound on $g_E(\mathbf X)$
yields the following concave lower-bound: $\tilde f^{(i)}(\mathbf X)
\triangleq
\log_2\bigl(1+\hat\gamma_B(\mathbf X)\bigr)
-
\bar g_E^{(i)}(\mathbf X)$. %which is tight at \(\mathbf X=\mathbf X^{(i)}\). 
%The first term in \eqref{eq:mm_obj} is concave in \(\mathbf X\), whereas \(\bar g_E^{(i)}(\mathbf X)\) is affine. Hence, \(\tilde f^{(i)}(\mathbf X)\) is concave and constitutes a global lower bound on the original secrecy objective that is tight at \(\mathbf X=\mathbf X^{(i)}\).
%Consequently, for fixed \(\{y_k\}\) and the current \(\mathbf X^{(i)}\), the reflection-coefficient subproblem is approximated by the following convex problem:
%\begin{equation}
%\label{prob:inner_sdp}
%\begin{aligned}
%\max_{\mathbf X\succeq\mathbf 0}
%~&
%\tilde f^{(i)}(\mathbf X)
%\\
%\text{s.t.}~&
%[\mathbf X]_{ii}\le g_{\max}^2~ (i=1,\cdots,M_o),
%\\
%&
%[\mathbf X]_{M_o+1,M_o+1}=1,
%\\
%&
%\operatorname{tr}\left(
%\tilde{\mathbf F}(\mathbf w)\mathbf X
%\right)
%\le P_{\max}.
%\end{aligned}
%\end{equation}
Consequently, for fixed $\{y_k\}$ and current $\mathbf X^{(i)}$,~\eqref{prob:P3X} is approximated by the following convex problem:
\begin{equation}
\label{prob:inner_sdp}
\begin{aligned}
\max_{\mathbf X\succeq0}
&~
\tilde f^{(i)}(\mathbf X) \\
\text{s.t.}~
&
\forall[\mathbf X]_{ii}\le g_{\max}^2, [\mathbf X]_{M_o+1,M_o+1}=1, \mathrm{tr}(\tilde{\mathbf F}(\mathbf w)\mathbf X)\le P_{\max}.
\end{aligned}
\end{equation}
%The above objective is still nonconcave due to the difference-of-concave structure. To obtain a tractable surrogate, we apply a majorization-minimization (MM) to the Eve term. XXXXThe first-order expansion of $\log_2 (1+x)$ at the current AO iterate $\mathbf X^{(i)}$ provides the following global upper-bound:
%%HJM 아래 수식 RHS에서 \hat{\gamma_E}(X)가 여전히 sqrt(u_E(X))를 포함해서 concave인거 같은데, 그래서 얘도 taylor로 한번 더 linear approximation 해줘야 되는게 아닌가 싶어서 한번 확인 부탁드립니다
%\begin{equation}
%\log_2(1+\hat\gamma_E(\mathbf X))
%\le
%\log_2(1+\hat\gamma_E^{(i)})
%+
%\frac{\hat\gamma_E(\mathbf X)-\hat\gamma_E^{(i)}}{(1+\hat\gamma_E^{(i)})\ln2},
%\label{eq:mm_upper}
%\end{equation}
%where $\hat\gamma_E^{(i)}\triangleq \hat\gamma_E(\mathbf X^{(i)})$. Substituting~\eqref{eq:mm_upper} into the secrecy objective yields the following concave lower-bound function:
%\begin{equation}
%\tilde f^{(i)}(\mathbf X)
%=
%\log_2(1+\hat\gamma_B(\mathbf X))
%-
%\left[
%\log_2(1+\hat\gamma_E^{(i)})
%+
%\frac{\hat\gamma_E(\mathbf X)-\hat\gamma_E^{(i)}}{(1+\hat\gamma_E^{(i)})\ln2}
%\right].
%\label{eq:mm_obj}
%\end{equation}
%Consequently, for fixed $\{y_k\}$ and current iterate $\mathbf X^{(i)}$,~\eqref{prob:P3X} is approximated by the following convex problem:
%\begin{equation}
%\label{prob:inner_sdp}
%\begin{aligned}
%\max_{\mathbf X\succeq0}
%&~
%\tilde f^{(i)}(\mathbf X) \\
%\text{s.t.}~
%&
%\forall[\mathbf X]_{ii}\le g_{\max}^2, [\mathbf X]_{M_o+1,M_o+1}=1, \mathrm{tr}(\tilde{\mathbf F}(\mathbf w)\mathbf X)\le P_{\max}.
%\end{aligned}
%\end{equation}
After obtaining the relaxed solution $\mathbf X^\star$, a feasible active reflection vector $\mathbf v^\star$ is recovered via Gaussian randomization~\cite{sdr}. Finally, the auxiliary variables are updated according to $y_k^{(i+1)}\leftarrow\frac{\sqrt{u_k(\mathbf X^\star)}}{v_k(\mathbf X^\star)}$, and refresh the linearization point in~\eqref{eq:mm_upper} by $\hat{\gamma}_E^{(i+1)}\leftarrow \hat{\gamma}_E^{\star}(\mathbf X^\star)$. %Since \(\tilde f^{(i)}(\mathbf X)\) constitutes a global concave lower-bound on the fixed-\(\{y_k\}\) log-difference objective and is tight at \(\mathbf X=\mathbf X^{(i)}\), the successive MM updates generate a non-decreasing sequence of objective values for the corresponding relaxed subproblem until convergence. %Since the objective is upper-bounded over the feasible set, the inner MM procedure converges. After its convergence, the auxiliary variables are updated in closed form, and the AO procedure continues until the actual secrecy objective satisfies a prescribed convergence tolerance.
Since~\eqref{eq:mm_upper} is a global upper-bound that is tight at \(\mathbf X=\mathbf X^{(i)}\), the MM procedure guarantees that the original log-difference objective is non-decreasing. Hence by alternately updating $\mathbf X(\leftrightarrow\mathbf v)$, $\hat{\gamma}_E$, and $\{y_k\}$, the objective is guaranteed to be non-decreasing until convergence.

\subsection{Update of $\mathbf S_{M_o}$: FARIS Port Configuration}
\label{subsubsec:CEO_rig}
Given $\mathbf w$ and $\mathbf v$, we optimize $\mathbf S_{M_o}$ via the cross-entropy optimization (CEO) method~\cite{FRISsec, FRISonoff}. Let a port-selection configuration be $\Gamma$ with $\Gamma \subseteq\{1,\cdots,M\}$ and $|\Gamma|=M_o$.
We parameterize a sampling distribution by a probability mass function (PMF) $\mathbf p^{(t-1)}=[p_1^{(t-1)} \cdots p_M^{(t-1)}]^{\mathrm T}$ with CEO iteration $t$. We then generate $K$ independent and identically distributed (i.i.d.) $\{\Gamma_n\}_{n=1}^K$ by drawing $M_o$ indices without replacement according to $\mathbf p^{(t-1)}$. For each $\Gamma_n$, we form $\mathbf S_{M_o}(\Gamma_n)$ and evaluate $
O(\Gamma_n)=\frac{1+\gamma_B(\mathbf w,\mathbf v,\mathbf S_{M_o}(\Gamma_n))}
{1+\gamma_E(\mathbf w,\mathbf v,\mathbf S_{M_o}(\Gamma_n))}$. After sorting $\{O(\Gamma_n)\}$, the top $K_e=\lceil\rho K\rceil$ samples with maximum $O(\Gamma_n)$ form the elite set $\mathcal E$. Thereafter, ${\mathbf p}^{(t)}$ is updated via $ p_m^{(t)}=\frac{1}{K_e M_o}\sum_{n:\Gamma_n\in\mathcal E}\mathbb I(m\in \Gamma_n)~( m=1,\cdots,M)$, followed by smoothing $\mathbf p^{(t)}\leftarrow(1-\alpha)\mathbf p^{(t-1)}+\alpha {\mathbf p}^{(t)}$ with smoothing factor $\alpha\in(0,1]$. The CEO iterations stop when $\{\mathbf p^{(t)}\}$ converges, and we select the indices corresponding to the $M_o$ largest probabilities to form $\mathbf S_{M_o}^\star$.
\subsection{Overall Algorithm and Complexity Analysis}
\label{subsec:convergence}
The proposed framework is summarized in Algorithm~\ref{alg:ao_ceo_faris}. Herein, the objective value is non-decreasing in each iteration~\cite{fracp, FRISonoff}, guaranteeing the convergence of the algorithm to a stationary point. For computational complexity, updating $\mathbf w$ requires solving the convex SDP in~\eqref{prob:P2_CC_rank} with an $N\times N$ semidefinite variable $\mathbf X$. Hence, the computational complexity grows on the order of $\mathcal O(I_{\rm SDP}N^{4.5})$~\cite{sdr}, where $I_{\rm SDP}$ denotes the number of solver iterations. If Gaussian randomization is used, an additional complexity of $\mathcal O(N_{\rm rand}N^{2})$ is incurred~\cite{sdr}. Updating $\mathbf v$ involves solving the lifted convex problem in~\eqref{prob:inner_sdp} with an $(M_o+1)\times(M_o+1)$ semidefinite variable $\mathbf X$. Similarly, the computational complexity scales on the order of $\mathcal O(I_{\rm cvx}(M_o+1)^{4.5})$, where $I_{\rm cvx}$ denotes the number of iterations required by the convex solver, with subsequent Gaussian randomization with $N_{\rm rand}'$ candidates introduces an additional $\mathcal O(N_{\rm rand}'M_o^{2})$. Updating $\mathbf S_{M_o}$ employs the CEO with $I_{\rm CEO}$ iterations and $K$ samples per iteration. Each sample evaluation requires computing the effective channels and amplified-noise terms, which can be implemented with complexity $\mathcal O(M_oN+M_o^{2})$. Hence, the total CEO complexity is $\mathcal O(I_{\rm CEO}K(M_oN+M_o^{2}))$. Overall, the total computational complexity over $I_{\rm AO}$ AO iterations can be approximated as
\begin{equation}
\begin{aligned}
\mathcal O\Big(
&I_{\rm AO}\big(
I_{\rm SDP}N^{4.5}
+N_{\rm rand}N^{2}\\
&+I_{\rm cvx}M_o^{4.5}
+N_{\rm rand}'M_o^{2}
+I_{\rm CEO}K(M_oN+M_o^{2})
\big)
\Big).
\end{aligned}
\end{equation}

%For computational complexity, updating $\mathbf w$ requires solving the convex SDP from~\eqref{prob:P2_CC_rank} with an $N\times N$ semidefinite $\mathbf X$. Using an interior-point method, the complexity is typically on the order of $\mathcal O(I_{\rm SDP}N^{4.5})$ with $I_{\rm SDP}$ iterations~\cite{sdr}. If Gaussian randomization is needed for rank-one recovery, an additional complexity of $\mathcal O(N_{\rm rand}N^2)$ is incurred~\cite{sdr}. Updating $\mathbf v$ also solves the lifted problem in~\eqref{prob:inner_sdp} with an $(M_o+1)\times(M_o+1)$-sized semidefinite $\mathbf X$; it is typically given by ${\mathcal O}(I_{\rm SDP}'M_o^{4.5})$ with $I_{\rm SDP}'$ iterations~\cite{boyd}, and the subsequent Gaussian Randomization with $N_{\rm rand}'$ candidates adds $\mathcal O(N_{\rm rand}' M_o^2)$~\cite{sdr}. Updating $\mathbf S_{M_o}$ runs CEO for $I_{\rm CEO}$ iterations with $K$ samples per iteration; each evaluation of $O(\Gamma_n)$ mainly requires forming the effective channels/noise terms, which can be implemented in $\mathcal O(M_oN+M_o^2)$, yielding a cost of $\mathcal O(I_{\rm CEO}K(M_oN+M_o^2))$. Therefore, the overall complexity over $I_{\rm AO}$ AO iterations is $\mathcal O\Big(I_{\rm AO}\big(I_{\rm SDP}N^{4.5} + N_{\rm rand}N^2+I_{\rm SDP}'M_o^{4.5} + N_{\rm rand}'M_o^2 + I_{\rm CEO}K(M_oN+M_o^2)\big)\Big)$.

\begin{algorithm}[t]
\caption{FARIS-Aided Secrecy Maximization}
\label{alg:ao_ceo_faris}
\begin{algorithmic}[1]
\State Initialize $(\mathbf v^{(0)},\mathbf S_{M_o}^{(0)}, \{y_k^{(0)}\}, \hat{\gamma}_E^{(0)})$ and set $i=0$.
%\State Initialize $(\mathbf v^{(0)},\mathbf S_{M_o}^{(0)})$ and set $i=0$.
\Repeat
\State $i\leftarrow i+1$.
\State Given $(\mathbf v^{(i-1)},\mathbf S_{M_o}^{(i-1)})$, solve~\eqref{prob:P2_exact} to obtain $\mathbf w^{(i)}$.
\State Given $(\mathbf w^{(i)},\mathbf S_{M_o}^{(i-1)})$, solve~\eqref{prob:inner_sdp} and recover $\mathbf v^{(i)}$.
\State Update $\{y_k^{(i)}\}$ and $\hat\gamma_E^{(i)}$.% by~\eqref{eq:y_update_mm} and~\eqref{eq:te_update_mm}, respectively.
\State Given $(\mathbf w^{(i)},\mathbf v^{(i)})$, run CEO to obtain $\mathbf S_{M_o}^{(i)}$.
\Until the objective value converges
\State Output: $(\mathbf w^\star,\mathbf v^\star,\mathbf S_{M_o}^\star)\leftarrow(\mathbf w^{(i)},\mathbf v^{(i)},\mathbf S_{M_o}^{(i)})$.
\end{algorithmic}
\end{algorithm}
\section{Simulation Results}
Unless stated otherwise, the AP with $N=4$, Bob, and Eve are located at $(0,0,10)$~m, $(50,0,2)$~m, and $(55,5,2)$~m, respectively. The FARIS with $M=64$, $M_o=25$, and $g_{\max}=40$~dB~\cite{aris5} is centered at $(45,10,5)$~m. The AP-FARIS channel follows a Rician fading model with exponent 2.2 and $K$-factor of $5$~dB, while all other links are subject to Rayleigh fading with exponent 2.8. To reflect a harsh non-line-of-sight (NLoS) environment suitable for FARIS/FRIS, we further impose a 25~dB blockage loss on the direct AP-user paths~\cite{rapalos, FRISsec}. The noise powers are set to $\sigma_r^2=\sigma_0^2=-90$~dBm, and the AP transmit and FARIS power constraint are given by $P_{\mathrm{AP}}=P_{\max,t}=25$~dBm with hardware powers of $(P_{\mathrm{DC}}, P_c)=(-5, -10)$~dBm. The parameters for CEO are chosen as $K=5M, \rho=0.15$ and $\alpha=0.7$~\cite{CEM}. %All results are averaged over $10^3$ independent Monte Carlo realizations.
For comparison, we consider the FRIS-assisted scheme in~\cite{FRISsec} as well as an AO-based benchmarks with passive/active-RIS (RIS/ARIS) and the scheme without reconfigurable surfaces.
\begin{figure}[t]
    \centering
    \subfloat[]{%
        \includegraphics[width=0.22\textwidth]{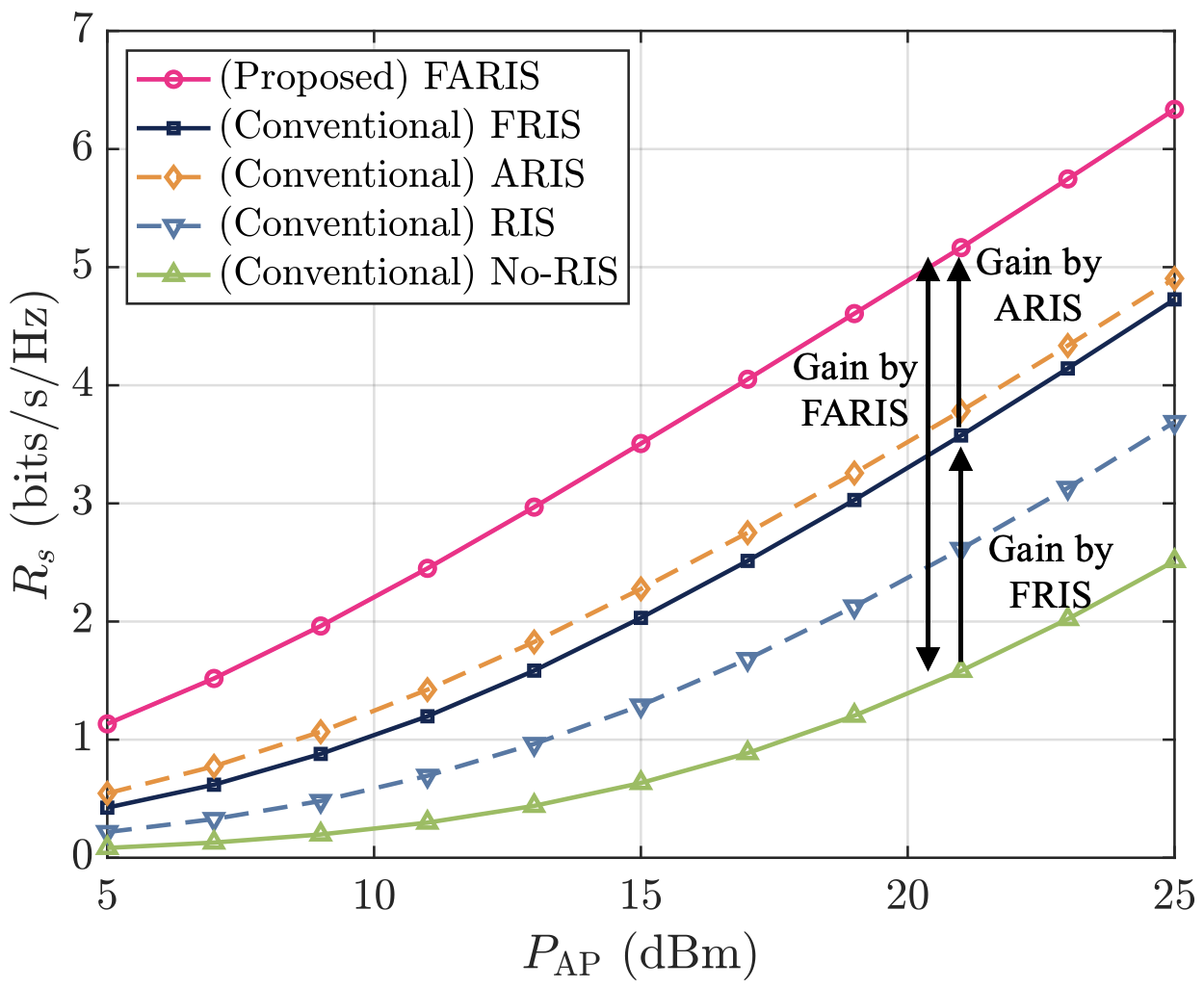}
        \label{fig_pap}%
    }
    \subfloat[]{%
        \includegraphics[width=0.22\textwidth]{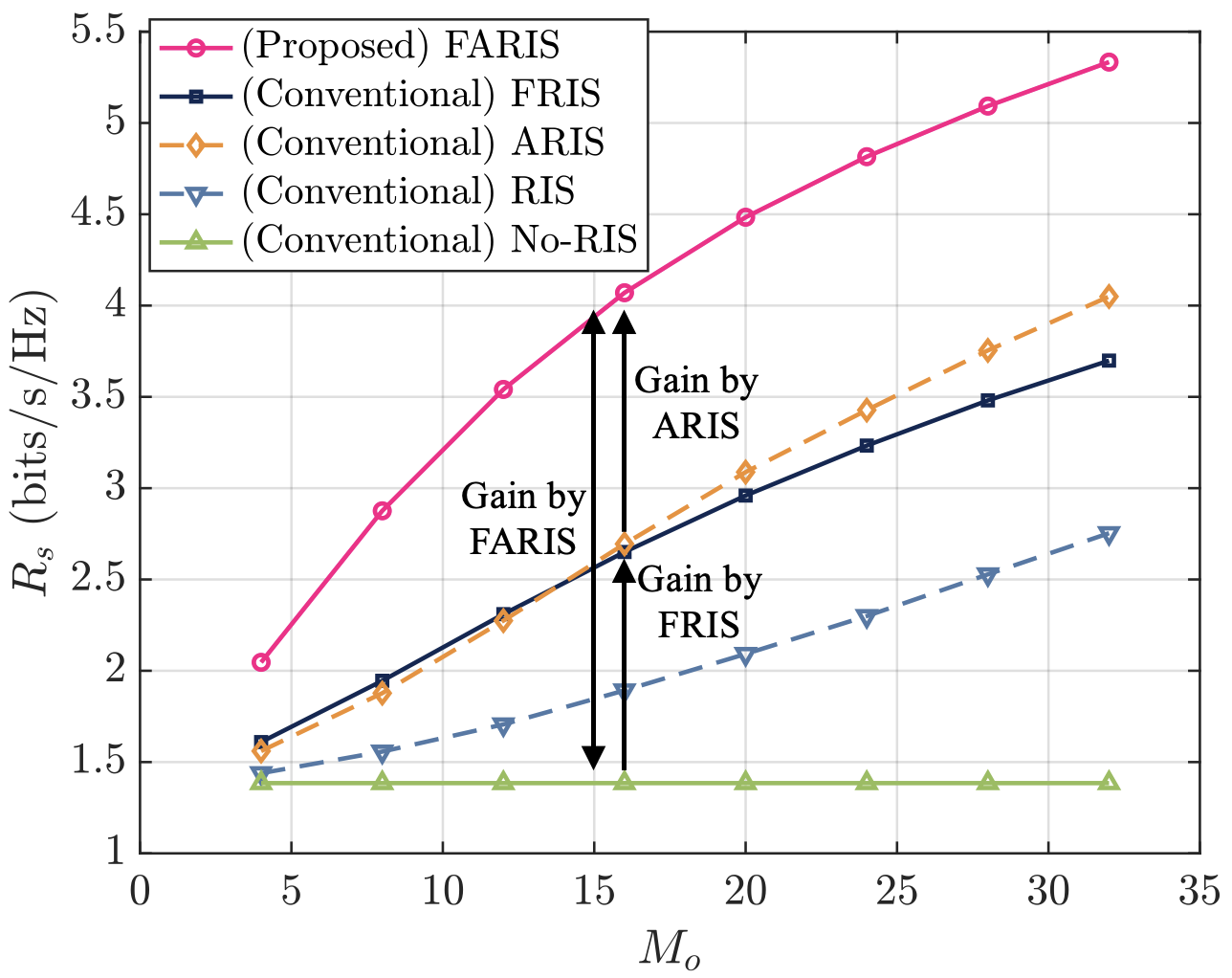}
        \label{fig_mo}%
    }
    \caption{Secrecy-rate according to (a) $P_{\mathrm{AP}}$ and (b) $M_o$.}
    \label{fig_1}
\end{figure}
%\begin{figure}[t]
%  \begin{center}
%    \includegraphics[width=0.5\columnwidth,keepaspectratio]{Fig/pap}
%    \caption{Secrecy-rate according to $P_{\mathrm{AP}}$.}
%    \label{fig_pap}
%  \end{center}
%\end{figure}

Fig.~\ref{fig_pap} shows the average secrecy rate as a function of $P_{\mathrm{AP}}$. As $P_{\mathrm{AP}}$ increases, all schemes achieve higher secrecy rates due to the improved received signal strength at Bob. However, the proposed FARIS consistently provides a significantly larger secrecy gain. This arises from the proposed joint optimization framework, which amplifies the disparity between the effective SINRs at Bob and Eve. %In contrast, the FRIS  lacks active gain control and therefore cannot fully exploit the increased power budget. The No-RIS scheme yields the lowest secrecy rate, underscoring the importance of FARIS/FRIS-assisted transmission in power-limited NLoS scenarios.
In contrast, the FRIS lacks active gain control and therefore cannot fully exploit the increased power budget. The ARIS/RIS exhibit intermediate performance: ARIS benefits from active amplification and slightly outperforms FRIS, whereas RIS trails FRIS because it lacks both active gain and fluid elements. The No-RIS scheme yields the lowest secrecy rate, underscoring the importance of RIS-assisted transmission in power-limited NLoS scenarios.

%\begin{figure}[t]
%  \begin{center}
%    \includegraphics[width=0.5\columnwidth,keepaspectratio]{Fig/mo}
%    \caption{Secrecy-rate according to $M_o$.}
%    \label{fig_mo}
%  \end{center}
%\end{figure}
Fig.~\ref{fig_mo} illustrates the average secrecy rate as a function of $M_o$. As $M_o$ increases, both FARIS- and FRIS-assisted schemes benefit from enhanced array and beamforming gains, resulting in a monotonic improvement in performance. Notably, the proposed FARIS consistently achieves the highest secrecy rate and exhibits a steeper growth trend than FRIS. This gain originates from the joint optimization of active reflection gains and port selection under a radiated-power constraint, which enables more effective shaping of the legitimate and eavesdropping channels. %In contrast, FRIS relies solely on passive phase adjustments, limiting its ability to fully exploit the additional spatial DoF introduced by increasing $M_o$. The No-RIS remains flat, confirming that increasing $M_o$ alone does not improve secrecy performance in the absence of reconfigurable surfaces.
In contrast, FRIS relies solely on passive nature, limiting its ability to fully exploit the additional spatial DoF by increased $M_o$. The ARIS/RIS also improve with $M_o$: ARIS increasingly benefits from active amplification and eventually outperforms FRIS, whereas RIS achieves more modest gains since it lacks both active gain and fluid elements, and trivially the No-RIS remains flat.%, confirming that increasing $M_o$ alone does not improve secrecy performance in the absence of reconfigurable surfaces.

\begin{figure}[t]
    \centering
    \subfloat[]{%
        \includegraphics[width=0.22\textwidth]{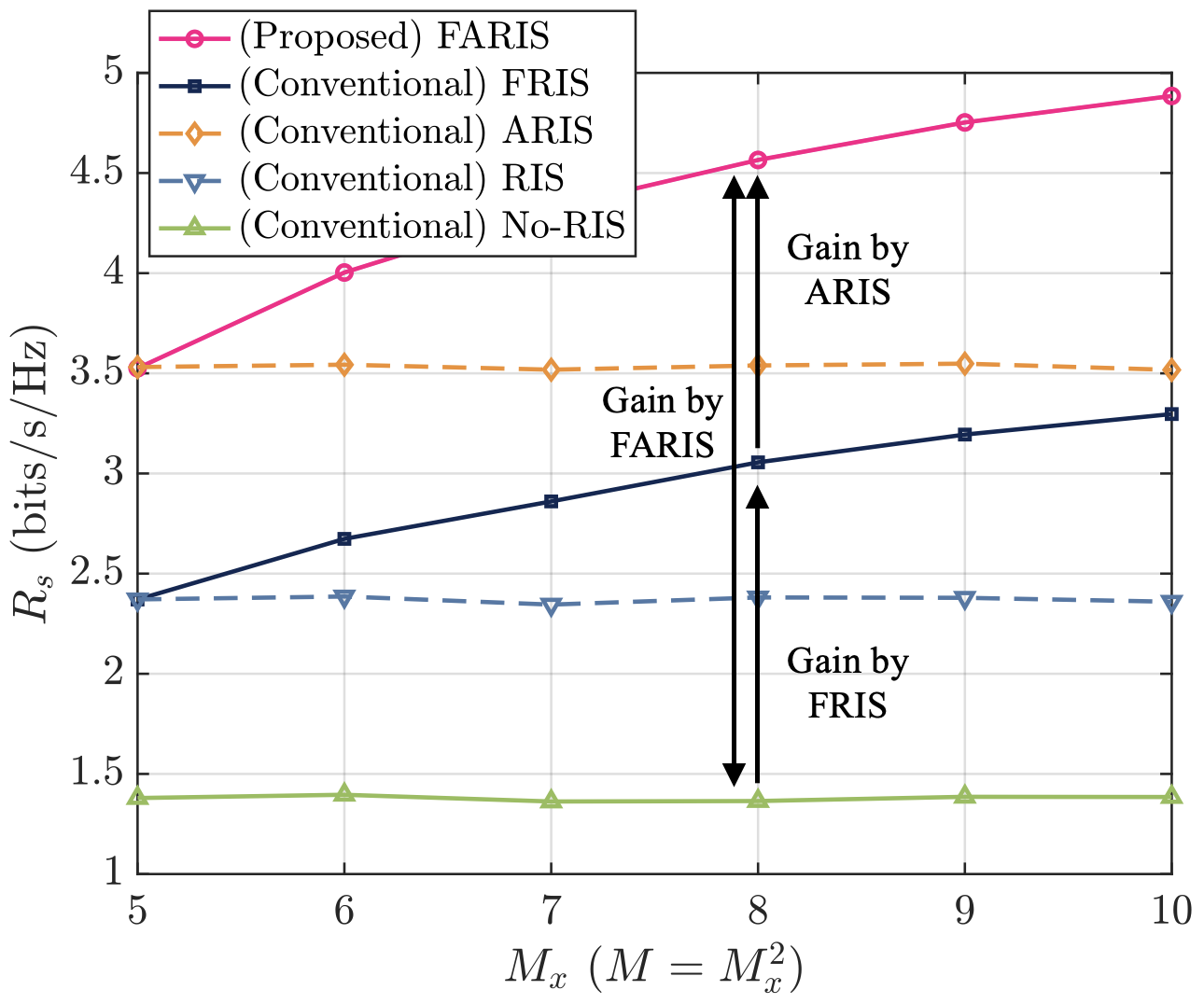}
        \label{fig_mx}%
    }
    \subfloat[]{%
        \includegraphics[width=0.22\textwidth]{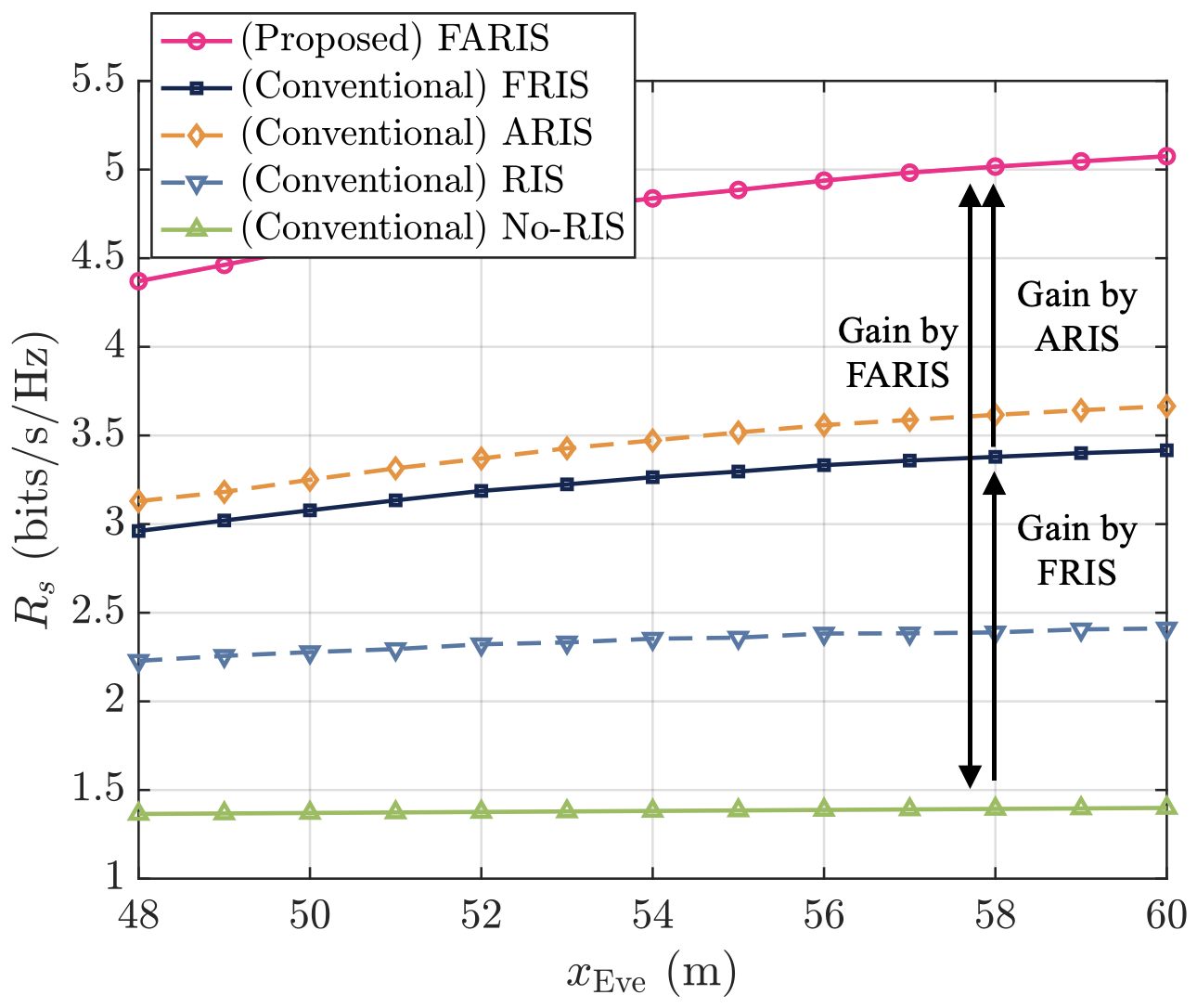}
        \label{fig_eve}%
    }
    \caption{Secrecy-rate according to (a) $M_x$ and (b) $x_{\mathrm{Eve}}$.}
    \label{fig_2}
\end{figure}
%\begin{figure}[t]
%  \begin{center}
%    \includegraphics[width=0.5\columnwidth,keepaspectratio]{Fig/mx}
%    \caption{Secrecy-rate according to $M_x$.}
%    \label{fig_mx}
%  \end{center}
%\end{figure} 
Fig.~\ref{fig_mx} investigates the effect by varying $M_x$. The secrecy rate of the proposed FARIS improves steadily with increasing $M_x$, highlighting the advantage of having a larger candidate set. This behavior reflects the effectiveness of the CEO-based port configuration, which exploits the denser candidate-port grid and finer spatial sampling available within the same physical aperture.%This behavior directly reflects the effectiveness of the CEO-based port configuration, which exploits the increased spatial diversity offered by a larger aperture.
%HJM M_x를 키우는게 port density를 올리고 effective aperture는 그대로인것 같아서 larger aperture 대신 port density 같이 수정하는게 좋을것 같습니다.
 %By comparison, the FRIS scheme exhibits a more moderate improvement since its passive architecture limits the achievable gain from aperture expansion. Note that the No-RIS scheme is insensitive to $M_x$, as it does not involve surface-assisted propagation. These results demonstrate that the integration of active reflection and fluid port selection is crucial for fully leveraging aperture growth.
By comparison, the FRIS scheme exhibits a more moderate improvement since its passive architecture limits the benefit obtainable from denser candidate-port selection. %gain from aperture expansion.
The ARIS/RIS remain nearly unchanged because they use a fixed $M_o$ elements and cannot exploit the additional candidate ports created by increasing $M_x$; nevertheless, ARIS maintains a higher secrecy rate than RIS owing to its active amplification capability. The No-RIS scheme is also insensitive to $M_x$, as it does not involve surface-assisted propagation. %These demonstrate that integrating active reflection with fluid port selection is crucial for fully leveraging aperture growth.

%\begin{figure}[t]
%  \begin{center}
%    \includegraphics[width=0.5\columnwidth,keepaspectratio]{Fig/xe}
%    \caption{Secrecy-rate according to $x_{\mathrm{Eve}}$.}
%    \label{fig_eve}
%  \end{center}
%\end{figure}
Fig.~\ref{fig_eve} evaluates the secrecy performance as a function of Eve’s horizontal position $x_{\mathrm{Eve}}$. Specifically, $x_{\mathrm{Eve}}$ is varied from $48$ to $60$~m, thereby changing Eve’s proximity to Bob located at $x=50$~m.As Eve moves closer to Bob, the secrecy rates of all schemes gradually decrease due to the reduced spatial separation and increased channel similarity~\cite{FRISsec}. Nevertheless, the proposed FARIS consistently maintains the highest secrecy rate across the entire range of Eve’s locations. This robustness stems from FARIS’s ability to adaptively reshape the spatial radiation pattern through active gain control and dynamic port selection, thereby mitigating information leakage even in unfavorable geometries. %The FRIS offers limited protection in such scenarios, while the No-RIS scheme is the most vulnerable. These results confirm that FARIS provides superior resilience against eavesdropping attacks in spatially challenging environments.
The ARIS provides stronger protection than FRIS owing to its active amplification capability, whereas RIS achieves a smaller but consistent secrecy improvement over the No-RIS scheme. Nevertheless, FRIS benefits from fluid port selection and outperforms RIS across the considered eavesdropper locations, while the No-RIS scheme remains the most vulnerable. %These results confirm that FARIS provides superior resilience against eavesdropping attacks in spatially challenging environments by jointly exploiting active reflection and fluid port selection.

\section{Conclusion}
In this paper, we demonstrated that FARIS fundamentally expands the design space of PLS by jointly leveraging active amplification and fluid port selection, enabling geometry-aware secrecy shaping beyond what RIS/FRIS architectures can achieve. A key insight was that secrecy enhancement is not driven solely by array gain, but by the ability of FARIS to selectively amplify and spatially reconfigure propagation paths under practical power constraints, thereby enlarging the SINR disparity between the legitimat user and the eavesdropper even in unfavorable geometries. The proposed framework revealed that FARIS operation requires holistic co-design of beamforming and element configuration while explicitly accounting for port correlation and amplification-induced noise. These findings confirm FARIS as a promising enabler for 6G PLS. %and motivate further investigation into the implementations to multi-user and wideband scenarios.
\bibliographystyle{IEEEtran}
\bibliography{IEEEexample}

\end{document}